\documentclass[showkeys,superscriptaddress,floatfix,prd,10pt,aps,twocolumn]{revtex4-2}
\usepackage{graphicx,epstopdf}
\usepackage[caption=false]{subfig}
\usepackage{appendix}
\usepackage[T1]{fontenc}
\usepackage{lmodern}
\usepackage[dvipsnames,x11names]{xcolor}
\usepackage[colorlinks=true,linkcolor=ForestGreen,citecolor=ForestGreen,urlcolor=NavyBlue]{hyperref}
\usepackage[sort&compress]{natbib}
\usepackage{lipsum}
\usepackage{morefloats}
\usepackage[pdf]{pstricks}
\usepackage{amsmath}
\usepackage{amssymb}
\usepackage{amsfonts}
\usepackage{rotating}
\usepackage{cancel}
\usepackage{mathtools}
\usepackage{bbm}
\usepackage{dsfont}
\usepackage{bbold}
\usepackage{multirow}
\usepackage{ulem}
\usepackage{physics}
\usepackage{orcidlink}
\usepackage{colortbl}

\def\pslash{p\!\!\!\slash }
\def\qslash{q\!\!\!\slash }
\def\xslash{x\!\!\!\slash }
\def\eslash{\varepsilon\!\!\!\slash }

\def\vel{\left|}
\def\ver{\right|}

\begin{document}

\title{Analysis of the $\Xi_c^* \bar K$ molecular pentaquark state by its electromagnetic properties}

\author{Ula\c{s}~\"{O}zdem\orcidlink{0000-0002-1907-2894}}%
\email[]{ulasozdem@aydin.edu.tr }
\affiliation{ Health Services Vocational School of Higher Education, Istanbul Aydin University, Sefakoy-Kucukcekmece, 34295 Istanbul, T\"{u}rkiye}

 
\begin{abstract}
We are systematically studying the electromagnetic characteristics of multiquark systems to shed light on their internal structure, whose nature and quantum numbers are controversial. In this study, we investigate the magnetic dipole, electric quadrupole, and magnetic octupole moments of the $\Xi_c^*\bar K$ state within the context of the QCD light-cone sum rule. During this analysis, we posit that the $\Xi_c^*\bar K$ state assumes a molecular structure with quantum numbers $J^P = \frac{3}{2}^-$. The extracted outcomes are given as  $\mu_{\Xi_c^*\bar K} = 0.15^{+0.04}_{-0.03}\,\mu_N$, $\mathcal{Q}_{\Xi_c^*\bar K} = (-0.93^{+0.22}_{-0.17})\times 10^{-3}\,\rm{fm^2}$, and $\mathcal{O}_{\Xi_c^*\bar K} = (-0.45^{+0.10}_{-0.09})\times 10^{-4}\,\rm{fm^3}$. The findings of this study, when considered alongside other pertinent characteristics, may assist in elucidating the nature of this controversial phenomenon.
\end{abstract}

\maketitle

\section{Motivation}
The members of the hadron sector, which contains a single heavy quark, are increasing in number and variety with each discovery.  In 2017, the LHCb collaboration conducted a mass spectrum investigation of the $\Xi_c^+ K^-$ channel. As a result of this investigation, the states $\Omega_c(3000)$, $\Omega_c(3050)$, $\Omega_c(3066)$, $\Omega_c(3090)$, and $\Omega_c(3119)$ were observed, as reported in Ref.~\cite{LHCb:2017uwr}. The preliminary analysis has yielded the conclusion that the five neutral resonances in question at least contain a heavy c-quark, in addition to two relatively light s-quarks. Later, the existence of the $\Omega_c(3000)$, $\Omega_c(3050)$, $\Omega_c(3066)$ and $\Omega_c(3090)$ states was validated by both the LHCb and Belle Collaborations through their observations~\cite{Belle:2017ext,LHCb:2023sxp}.  The observation of these states has prompted considerable research in the field of baryon spectroscopy, intending to elucidate their internal structure and potentially determining their hitherto unknown values of spin-parity in the context of different approaches and predictions~\cite{Chiladze:1997ev,Chen:2017xat,Nieves:2017jjx,Montana:2017kjw,Debastiani:2017ewu,Wang:2017vnc,Chen:2017gnu,Luo:2021dvj,Galkin:2020iat,Roberts:2007ni,Shah:2016mig,Yoshida:2015tia,Karliner:2017kfm,Wang:2017hej,Agaev:2017lip,Azizi:2018dva,Santopinto:2018ljf,Kim:2017jpx,An:2017lwg,Huang:2017dwn,Wang:2017zjw,Chen:2017sci,Chen:2015kpa, Padmanath:2017lng,Cheng:2017ove,Zhao:2017fov,Zhu:2022fyb}. 
 
One of the most popular explanations is that the discovered states correspond to the excited states of the $\Omega_c$ baryons. However, the possibility of the pentaquark is not excluded and is studied in the literature.  For example, in Ref.~\cite{Kim:2017jpx}, the authors employed the chiral-quark soliton model to classify the $\Omega_c$ states recently reported by the LHCb collaboration. They identified the observed $\Omega_c(3000)$, $\Omega_c(3066)$ and $\Omega_c(3090)$ with $\left(J=0, 1/2^-\right)$ and $\left(J=1,   1/2^-,3/2^-\right)$ states from the excited sextet, whereas they identified the most narrow $\Omega_c(3050)$ and $\Omega_c(3119)$ states with the $\left(J=1, 1/2^+,3/2^+\right)$ states from {the} exotic $\overline{\mathbf 15}$ multiplet. In Ref.~\cite{Huang:2017dwn}, the authors studied the $\Omega_{c}^{0}$ states as the $S-$wave molecular pentaquarks with $I=0$, $J^{P}=\frac{1}{2}^{-}$, $\frac{3}{2}^{-}$, and $\frac{5}{2}^{-}$ by solving the RGM equation in the framework of chiral quark model. Their predictions suggest that $\Omega_{c}(3119)^{0}$ can be explained as an $S-$wave resonance state of $\Xi D$ with $J^{P}=\frac{1}{2}^{-}$, and the decay channels are the $S-$wave $\Xi_{c} K$ and $\Xi^{'}_{c}K$. Other observed $\Omega^{0}_{c}$ states cannot be extracted in the present analysis. In Ref.~\cite{An:2017lwg}, the spectrum of several low-lying $sscq\bar{q}$ pentaquark configurations are obtained using the constituent quark model. Their results imply that four $sscq\bar{q}$ configurations with $J^{P}=1/2^{-}$ or $J^{P}=3/2^{-}$ lie at energies very close to the $\Omega_{c}^{0}$ states observed by the LHCb collaboration. In Ref.~\cite{Montana:2017kjw}, authors have studied the interaction of the low-lying pseudoscalar mesons with the ground-state baryons in the charm $+1$, strangeness $-2$ and isospin-0 sector, employing a t-channel vector meson exchange model with effective Lagrangians. They conclude that two of ($\Omega_c(3050,3090)$) the five $\Omega_c$ states observed by the LHCb collaboration could have a meson-baryon molecular origin. In Ref.~\cite{Debastiani:2017ewu}, the authors have studied $\Omega_c$ states which are generated dynamically from the interaction in the meson-baryon in the charm sector using an extension of the local hidden gauge approach with the exchange of vector mesons.  They predicted that two states with $J^P = 1/2^-$, which are remarkably close in mass and width to the $\Omega_c(3050,3090)$  states.  In Ref.~\cite{Chen:2017xat}, the authors have performed a coupled channel analysis to search for possible $\Omega_c$-like molecular states by using a one-boson-exchange potential. According to their results, states $\Omega_c(3090)$ and $\Omega_c(3119)$ can be in molecular structure. In Ref. \cite{Nieves:2017jjx}, the authors have applied the renormalization procedure used in the unitarized coupled-channel model and its implications in the charm$=1$, strangeness$=-2$, and isospin$=0$ sectors, where five $\Omega_c$ states have recently been observed by the LHCb collaboration.  They obtained three baryon-meson molecular states that couple predominantly to $\bar K \Xi_c^{'}$, $D \Xi$ and $\bar K \Xi_c^*$, respectively, but with a different experimental mass assignment, i.e. $J=1/2$ $\Omega_c(3000)$ and $J=1/2$ $\Omega_c(3119)$ or $\Omega_c(3090)$ and $J=3/2$ $\Omega_c(3050)$, corresponding to poles {\bf c} and {\bf d}, and {\bf b}, respectively.  In Ref.~\cite{Azizi:2018dva}, the authors utilized the QCD sum rules to investigate the masses of the $\Omega_c$ states. Their predictions were found to be consistent with the experimental data within the errors. In Ref. \cite{Zhu:2022fyb}, the  $\Omega_c$-like molecular states have been explored in a quasipotential Bethe-Salpeter equation approach.  Their results suggest that an isoscalar state can be produced from the $\Xi^*_c\bar{K}$ interaction with spin parity $3/2^-$, which can be related to state $\Omega_c(3120)$. Its isoscalar partner is predicted with a dominant decay in the $\Omega_c^*\pi$ channel. The isoscalar and isovector states with $1/2^-$ can be produced from the $\Xi'_c\bar{K}$ interaction with a threshold close to the mass of the $\Omega_c(3050)$ and $\Omega_c(3065)$. In Refs.~\cite{Wang:2021cku,Wang:2018alb}, the author has applied the QCD sum rules method to study the mass of $\Omega_c$ states with $J^P =1/2^-(3/2^-)$ quantum numbers and assumed that these states have diquark-diquark-antiquark structure. According to the author's predictions, the masses of these states are compatible with the experimental discoveries within errors.

A review of the literature reveals that the majority of research has concentrated on calculating the spectroscopic parameters of the $\Omega_c$ states. However, relying solely on spectroscopic parameters is inadequate to resolve the controversial nature of these states. Consequently, to determine the precise internal configurations of these states, additional studies are required, including the acquisition of electromagnetic multipole moments, and the investigation of radiative decays and weak decays. In order to gain insight into the nature and internal structure of controversial states, it is essential to consider the physical quantities that are related to the electromagnetic properties of hadrons. The distributions of charge and magnetism within hadrons can be investigated through the use of electromagnetic form factors and multipole moments, which arise from the zero-momentum transfer of these form factors. This information may be employed to delineate the distribution of quark-antiquark pairs and gluons within the hadron. Furthermore, it offers insights into the geometric forms, electric radii, and magnetic radii of hadrons. 
 
As mentioned above, the electromagnetic form factors and the resultant multipole moments play a pivotal role in this field, as they represent a fundamental quantity that encapsulates the composition and dynamical properties of hadrons. Inspired by this perspective, in this study, we investigate the magnetic dipole, electric quadrupole, and magnetic octupole moments of the $\Xi_c^*\bar K$ state.  
During this analysis, we posit that the $\Xi_c^*\bar K$ state assumes a molecular structure with quantum numbers $J^P = \frac{3}{2}^-$. To investigate the magnetic dipole, electric quadrupole, and magnetic octupole moments of hadrons, which are the parameters that belong to the low-energy regime of the QCD, it is necessary to employ reliable and efficient non-perturbative methods. One such method is the QCD light-cone sum rules method~\cite{Chernyak:1990ag,Braun:1988qv,Balitsky:1989ry}, which has been highly successful in predicting the static and dynamical properties of the hadron sector. In this approach, the physical parameters of the related hadrons are extracted by calculating the correlation function at both the hadron and the QCD levels. Subsequently, the Borel transform and continuum subtraction routines are employed to attenuate the impact of continuum and higher states effects. In conjunction with these aforementioned processes, dispersion integrals and the assumption of quark-hadron duality have also been implemented. Several studies in the literature have examined the electromagnetic characteristics of the singly-heavy hadrons that are currently the subject of controversy~\cite{Ozdem:2024pyb,Ozdem:2023edw,Ozdem:2023okg,Ozdem:2023eyz,Ozdem:2022ydv,Ozdem:2021vry,Azizi:2021aib,Azizi:2018jky,Azizi:2018mte}.

The manuscript is organized in the following manner: In Sec.~\ref{secII}, we extract the electromagnetic characteristics of the $\Xi_c^*\bar K$ state within the QCD light-cone sum rules. The numerical results and discussions about the magnetic dipole, electric quadrupole, and magnetic octupole moments are presented in Sect. \ref{secIII}. This work ends with a summary in Sect.~\ref{secIV}.

 \begin{widetext}
  
\section{Formalism}\label{secII}

The analysis of the electromagnetic properties of the molecular pentaquark $\Xi_c^*\bar K$ in the context of QCD light-cone rules commences with the formulation of the correlation function as follows:
 \begin{align} \label{edmn01}
\Pi_{\mu\nu}(p,q)&=i\int d^4x e^{ip \cdot x} \langle0|T\left\{J_\mu(x)\bar{J}_\nu(0)\right\}|0\rangle _\gamma \,,
\end{align}
where $T$ is the time ordered product, $\gamma$  is the external electromagnetic field, and the $J_\mu(x)$ is the interpolating current for the $\Xi_c^*\bar K$ state. The explicit form is written as follows 
\begin{align}
 \label{curpcs1}
J_\mu(x)&= \mid \Xi_c^*(x) \bar K(x)  \rangle=\big[ \varepsilon^{abc} d^{a^T}(x)  C\gamma_\mu s^b(x) c^{c}(x)\big]\big[  \bar d^{d}(x)\gamma_5 s^d(x)\big],
\end{align}
where the $C$ is the charge conjugation operator and $a$, $b$... are color indices.

To determine the hadron level, a complete set of hadron states with identical quantum numbers is incorporated into the correlation function in conjunction with the aforementioned interpolating current.  Subsequently, integration over x is conducted, resulting in the following outcome,
 \begin{align}
\Pi^{Had}_{\mu\nu}(p,q)&=\frac{\langle0\mid  J_{\mu}(x)\mid
{\Xi_c^*\bar K}(p,s)\rangle}{[p^{2}-m_{{\Xi_c^*\bar K}}^{2}]}
\langle {\Xi_c^*\bar K}(p,s)\mid
{\Xi_c^*\bar K}(p+q,s)\rangle_\gamma
\frac{\langle {\Xi_c^*\bar K}(p+q,s)\mid
\bar{J}_{\nu}(0)\mid 0\rangle}{[(p+q)^{2}-m_{{\Xi_c^*\bar K}}^{2}]}\nonumber\\
&+\mbox{continuum and higher states}.\label{Pc103}
\end{align}
The matrix elements of Eq.~(\ref{Pc103}) contain hadronic parameters such as spinors ($u_{\mu}$), form factors ($F_i$), residues ($\lambda$) and so forth. These are as follows:
\begin{align}
\langle0\mid J_{\mu}(x)\mid {\Xi_c^*\bar K}(p,s)\rangle&=\lambda_{{\Xi_c^*\bar K}}u_{\mu}(p,s),\\
\langle {\Xi_c^*\bar K}(p+q,s)\mid\bar{J}_{\nu}(0)\mid 0\rangle &= \lambda_{{\Xi_c^*\bar K}}\bar u_{\nu}(p+q,s), \\
\langle {\Xi_c^*\bar K}(p,s)\mid {\Xi_c^*\bar K}(p+q,s)\rangle_\gamma &=-e\bar
u_{\mu}(p,s)\bigg[F_{1}(q^2)g_{\mu\nu}\eslash 
-
\frac{1}{2m_{{\Xi_c^*\bar K}}} 
\Big[F_{2}(q^2)g_{\mu\nu} 
+F_{4}(q^2)\frac{q_{\mu}q_{\nu}}{(2m_{{\Xi_c^*\bar K}})^2}\Big]\eslash\qslash
\nonumber\\
&+
F_{3}(q^2)\frac{1}{(2m_{{\Xi_c^*\bar K}})^2}q_{\mu}q_{\nu}\eslash \bigg] 
u_{\nu}(p+q,s).
\label{matelpar}
\end{align}
%
%

The formulas presented above, when applied with the requisite mathematical simplifications, yield
\begin{align}
\Pi^{Had}_{\mu\nu}(p,q)&=-\frac{\lambda_{\Xi_c^*\bar K}^{2}}{[(p+q)^{2}-m_{\Xi_c^*\bar K}^{2}][p^{2}-m_{\Xi_c^*\bar K}^{2}]}
\bigg[  - F_{1}(q^2)\,g_{\mu\nu}\pslash\eslash\qslash 
+F_{2}(q^2)\,m_{\Xi_c^*\bar K}g_{\mu\nu}\eslash\qslash+
\frac{F_{3}(q^2)}{4m_{\Xi_c^*\bar K}}q_{\mu}q_{\nu}\eslash\qslash\, \nonumber\\&+
\frac{F_{4}(q^2)}{4m_{\Xi_c^*\bar K}^3}(\varepsilon.p)q_{\mu}q_{\nu}\pslash\qslash \,+
\rm{other~independent~Lorentz~structures} \bigg].
\label{final phenpart}
\end{align}

As they are more accessible experimentally, we require the magnetic and higher multipole form factors as a function of the form factors $\rm{F_i}(q^2)$. They are given by the following: 
\begin{align}
\label{edmn07}
G_{M}(q^2) &= \left[ F_1(q^2) + F_2(q^2)\right] ( 1+ \frac{4}{5}
\tau ) -\frac{2}{5} \left[ F_3(q^2) 
 +
F_4(q^2)\right] \tau \left( 1 + \tau \right), \\
G_{Q}(q^2) &= \left[ F_1(q^2) -\tau F_2(q^2) \right]  -
\frac{1}{2}\left[ F_3(q^2) -\tau F_4(q^2)
 ( 1+ \tau \right),   \\
 G_{O}(q^2) &=
\left[ F_1(q^2) + F_2(q^2)\right] -\frac{1}{2} \left[ F_3(q^2)  +
F_4(q^2)\right] 
 \left
 ( 1 + \tau \right),
\end{align}  
where   $\tau
= -\frac{q^2}{4m^2_{{\Xi_c^*\bar K}}}$.  There is a limitation here, which is that we are working with a real photon, which forces us to work in zero momentum transfer. Therefore, the form factors that have been defined above are formulated as magnetic dipole moments $(\mu)$, electric quadrupole $(Q)$ and magnetic octupole $(O)$ moments at zero momentum transfer as follows: 
 \begin{align}\label{mqo2}
\mu_{\Xi_c^*\bar K}&=\frac{e}{2m_{\Xi_c^*\bar K}}G_{M}(0),\\
Q_{\Xi_c^*\bar K}&=\frac{e}{m_{\Xi_c^*\bar K}^2}G_{Q}(0),\\
O_{\Xi_c^*\bar K}&=\frac{e}{2m_{\Xi_c^*\bar K}^3}G_{O}(0), 
\end{align}
with $G_{M}(0)$, $G_{Q}(0)$ and $G_{O}(0)$ being
 
\begin{align}\label{mqo1}
G_{M}(0)&=F_{1}(0)+F_{2}(0),\\
G_{Q}(0)&=F_{1}(0)-\frac{1}{2}F_{3}(0),\\
G_{O}(0)&=F_{1}(0)+F_{2}(0)-\frac{1}{2}[F_{3}(0)+F_{4}(0)].
\end{align}

Having completed the calculations at the hadron level, it is now necessary to proceed to the quark-gluon level.  The magnetic dipole, electric quadrupole, and magnetic octupole moments have been calculated at the quark-gluon level, employing the interpolating current in the correlation function.   With the help of Wick's theorem, all corresponding quark fields are contracted and the following equality is obtained: 
\begin{align}
\Pi^{\rm{QCD}}_{\mu\nu}(p,q)=&-i\,\varepsilon^{abc}\varepsilon^{a^{\prime}b^{\prime}c^{\prime}}\int d^4x e^{ip\cdot x} \langle 0| 
\bigg\{
\rm{Tr}\Big[\gamma_5  S_s^{dd^\prime}(x) \gamma_5  S_d^{d^\prime d}(-x)\Big]
\rm{Tr}\Big[\gamma_\mu  S_{s}^{bb^\prime}(x) \gamma_\nu \widetilde S_{d}^{aa^\prime}(x)\Big] S_c^{cc^\prime}(x) \nonumber\\
&-
\rm{Tr}\Big[\gamma_5  S_s^{db^\prime}(x) \gamma_\nu  \widetilde S_{d}^{aa^\prime}(x)  
\gamma_\mu  S_{s}^{bd^\prime}(x) \gamma_5  S_{d}^{d^\prime d}(-x)\Big] S_c^{cc^\prime} (x) \bigg\} 
|0 \rangle_\gamma,\label{QCD1}
\end{align}
where   
$\widetilde{S}_{c(q)}^{ij}(x)=CS_{c(q)}^{ij\rm{T}}(x)C$.   The corresponding quark propagators are defined as follows~\cite{Yang:1993bp, Belyaev:1985wza}: 
\begin{align}
\label{edmn13}
S_{q}(x)&= S_q^{free}(x) 
- \frac{\langle \bar qq \rangle }{12} \Big(1-i\frac{m_{q} \xslash}{4}   \Big)
- \frac{ \langle \bar qq \rangle }{192}
m_0^2 x^2  \Big(1 
  -i\frac{m_{q} \xslash}{6}   \Big)
+\frac {i g_s~G^{\mu \nu} (x)}{32 \pi^2 x^2} 
\Big[\rlap/{x} 
\sigma_{\mu \nu} +  \sigma_{\mu \nu} \rlap/{x}
 \Big],\\
%
S_{c}(x)&=S_c^{free}(x)
-\frac{m_{c}\,g_{s}\, G^{\mu \nu}(x)}{32\pi ^{2}} \bigg[ (\sigma _{\mu \nu }{\xslash}
+{\xslash}\sigma _{\mu \nu }) 
    \frac{K_{1}\big( m_{c}\sqrt{-x^{2}}\big) }{\sqrt{-x^{2}}}
 +2\sigma_{\mu \nu }K_{0}\big( m_{c}\sqrt{-x^{2}}\big)\bigg],
 \label{edmn14}
\end{align}%
with  
\begin{align}
 S_q^{free}(x)&=\frac{1}{2 \pi x^2}\Big(i \frac{\xslash}{x^2}- \frac{m_q}{2}\Big),\\
 S_c^{free}(x)&=\frac{m_{c}^{2}}{4 \pi^{2}} \bigg[ \frac{K_{1}\big(m_{c}\sqrt{-x^{2}}\big) }{\sqrt{-x^{2}}}
+i\frac{{\xslash}~K_{2}\big( m_{c}\sqrt{-x^{2}}\big)}
{(\sqrt{-x^{2}})^{2}}\bigg].
\end{align}
%

To conduct further calculations at the quark-gluon level, it is essential to take into account two different contributions arising from the short- and long-distance interactions of the photon with quarks. In order to determine the nature of the short-distance interactions, the subsequent replacement must be carried out following the procedure outlined below.
\begin{align}
\label{free}
S^{free}(x) \rightarrow \int d^4z\, S^{free} (x-z)\,\rlap/{\!A}(z)\, S^{free} (z)\,.
\end{align}
In this scheme, one of the propagators for a light or heavy quark interacts with the photon at a short-distance, while the remaining four propagators are assumed to be free.

The following equation will be employed to incorporate long-distance contributions into the analysis. 
 \begin{align}
\label{edmn21}
S_{\alpha\beta}^{ab}(x) \rightarrow -\frac{1}{4} \big[\bar{q}^a(x) \Gamma_i q^b(0)\big]\big(\Gamma_i\big)_{\alpha\beta},
\end{align}
where the four remaining propagators are supposed to be full ones. Here $\Gamma_i = \{\textbf{1}$, $\gamma_5$, $\gamma_\mu$, $i\gamma_5 \gamma_\mu$, $\sigma_{\mu\nu}/2\}$. Inserting Eq.~(\ref{edmn21}) into Eq.~(\ref{QCD1}) yields  $\langle \gamma(q)\vel \bar{q}(x) \Gamma_i q(0) \ver 0\rangle$ and $\langle \gamma(q)\vel \bar{q}(x) \Gamma_i G_{\alpha\beta}q(0) \ver 0\rangle$  matrix elements, which are defined concerning the photon distribution amplitudes (DAs) and denote the interaction of photons with quark fields at large distances.  The aforementioned matrix elements play a pivotal role in the continued analysis of non-perturbative contributions~(See Ref.~\cite{Ball:2002ps} for details on the DAs of the photon).   Following the completion of the aforementioned procedures, the requisite electromagnetic multipole moments, calculated at the quark-gluon level for the specified state, have been duly evaluated. Further details regarding the methodology employed to derive the short- and long-distance contributions can be seen in~\cite{Ozdem:2022eds,Ozdem:2022vip}.

Analytical expressions for the electromagnetic multipole moments of the $\Xi_c^* \bar K$ state have been derived at both the hadron and quark-gluon levels. As a final step, by matching the coefficients of the  $g_{\mu\nu}\pslash\eslash\qslash$, $g_{\mu\nu}\eslash\qslash$, $q_{\mu}q_{\nu}\eslash\qslash$ and  $(\varepsilon.p)q_{\mu}q_{\nu}\pslash\qslash$  structures, respectively, for the $F_1(0)$, $F_2(0)$, $F_3(0)$ and $F_4(0)$ form factors, we get the corresponding sum rules for these form factors. The final forms of these form factors can be found in the Appendix.

\end{widetext}

\section{Numerical results and discussion} \label{secIII}

The purpose of this section is to perform a numerical analysis of the magnetic dipole, electric quadrupole, and magnetic octupole moments of the $\Xi_c^* \bar K$ state. The following inputs are employed in the analysis of the physical quantities to be computed:  
  $m_s =93.4^{+8.6}_{-3.4}\,\mbox{MeV}$, $m_{\Xi_c^* \bar K} = 3049^{+155}_{-149}$~MeV~\cite{Azizi:2018dva},  
$m_c = 1.27\pm 0.02\,$GeV~\cite{ParticleDataGroup:2022pth}, $\langle \bar qq\rangle $=$(-0.24\pm 0.01)^3\,$GeV$^3$,  $\langle \bar ss\rangle $= $0.8 \langle \bar qq\rangle$~\cite{Ioffe:2005ym}, $\lambda_{\Xi_c^* \bar K} = (2.59^{+0.36}_{-0.36}) \times 10^{-4} \rm{GeV^{6}}$~\cite{Azizi:2018dva}, 
$m_0^{2} = 0.8 \pm 0.1$~GeV$^2$, $\langle g_s^2G^2\rangle = 0.48 \pm 0.14~ $GeV$^4$~\cite{Narison:2018nbv}, $\chi=-2.85 \pm 0.5~\mbox{GeV}^{-2}$~\cite{Rohrwild:2007yt}, and $f_{3\gamma}=-0.0039~$GeV$^2$~\cite{Ball:2002ps}. In numerical calculations, we set $m_u$ =$m_d$ = 0 and $m^2_s = 0$, but take into account terms proportional to $m_s$. The pertinent details regarding the photon DAs and the specific parameters utilized in these DAs are elucidated in \cite{Ball:2002ps}.

\begin{widetext}
 
 \begin{figure}[htp]
\centering
 \subfloat[]{\includegraphics[width=0.37\textwidth]{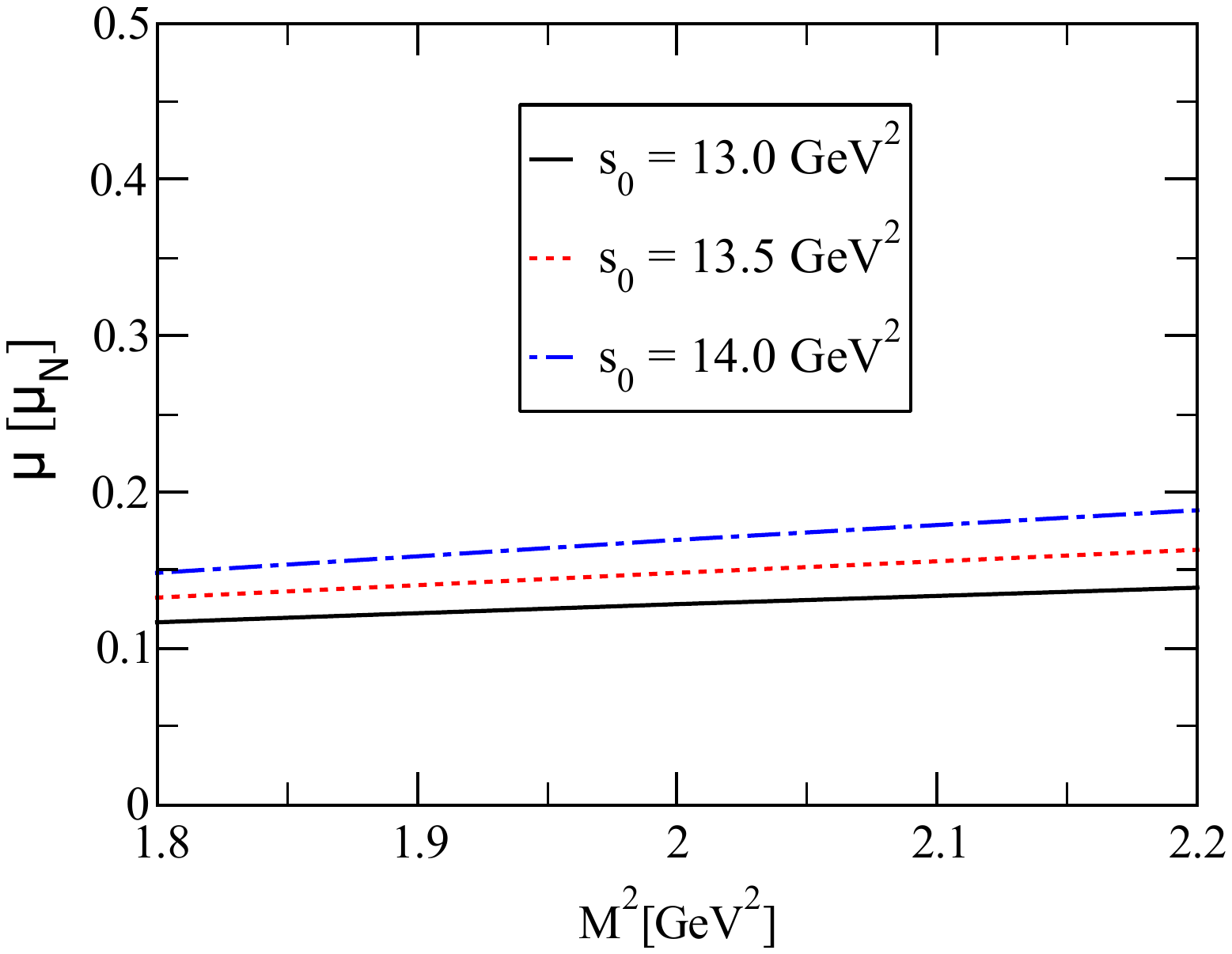}}~~~~~~~
 \subfloat[]{\includegraphics[width=0.37\textwidth]{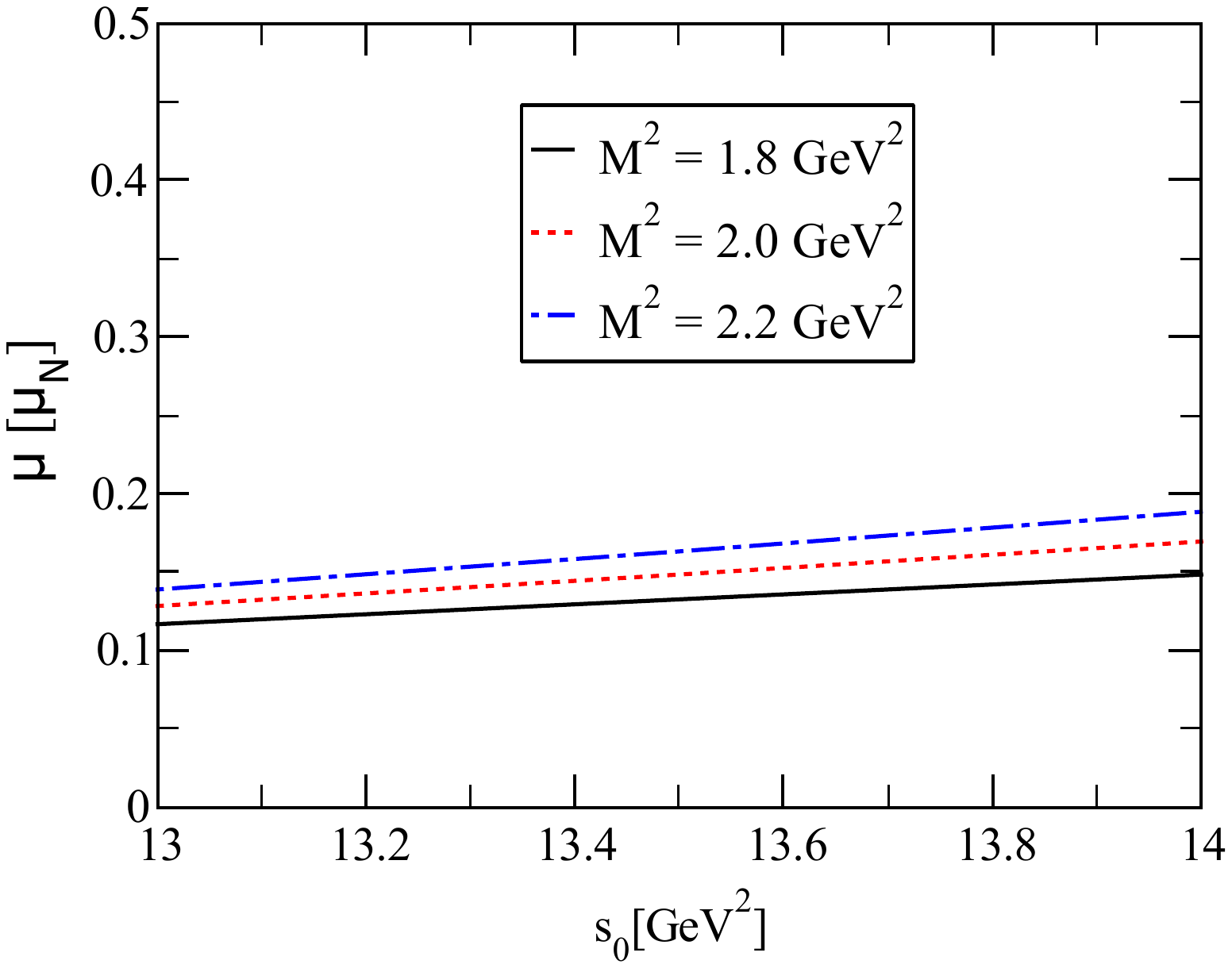}}\\
 \subfloat[]{ \includegraphics[width=0.37\textwidth]{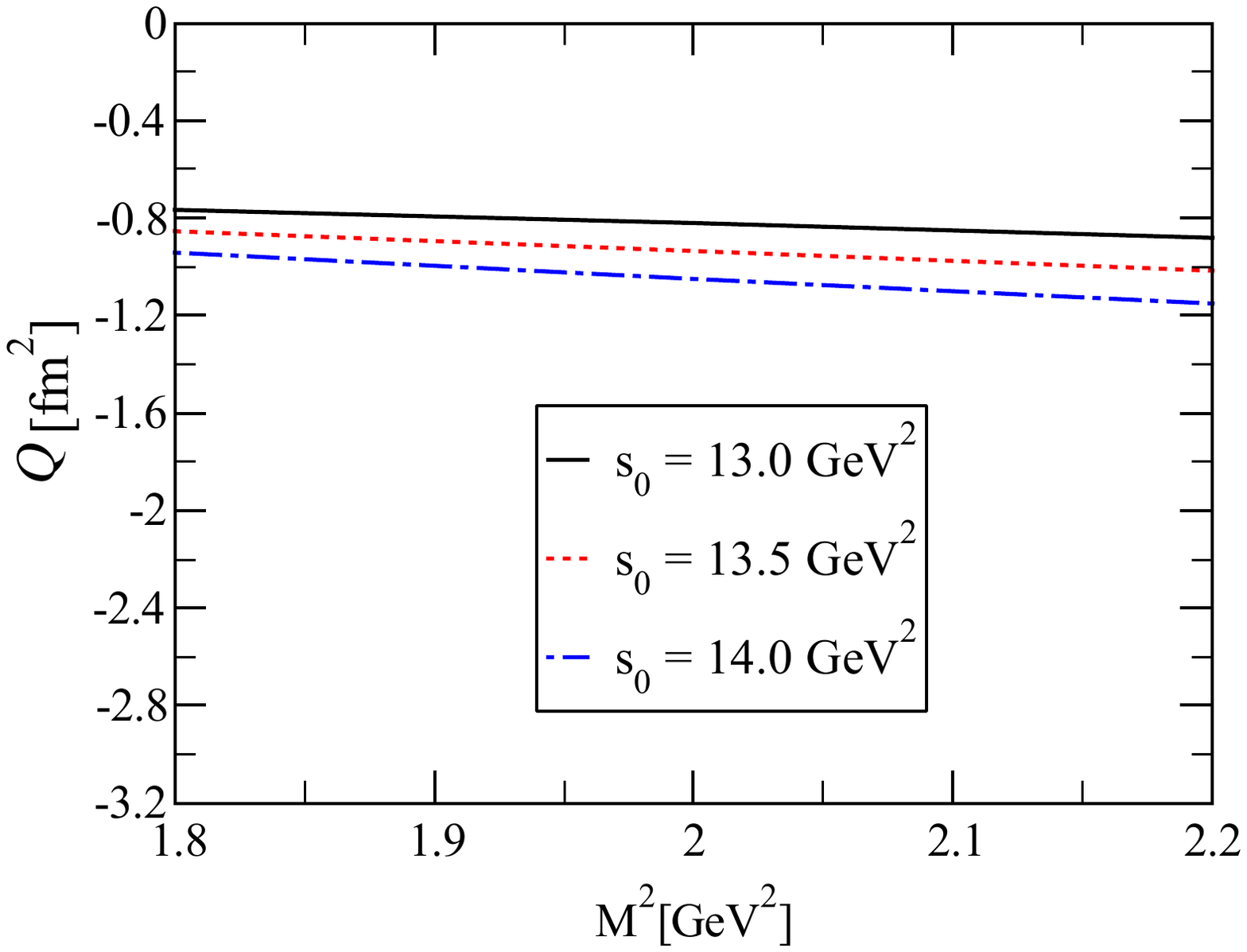}}~~~~~~~
 \subfloat[]{\includegraphics[width=0.37\textwidth]{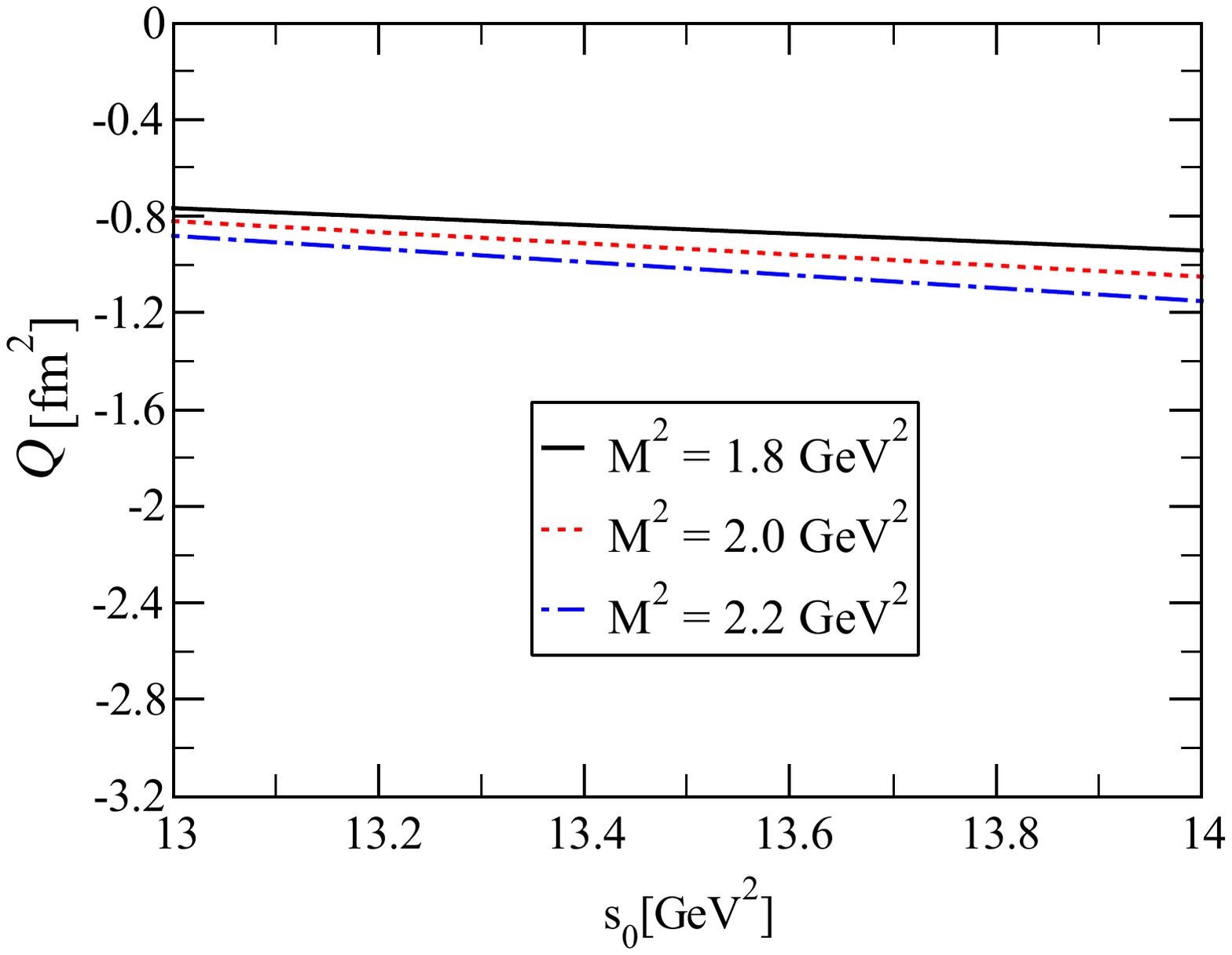}}\\
 \subfloat[]{ \includegraphics[width=0.37\textwidth]{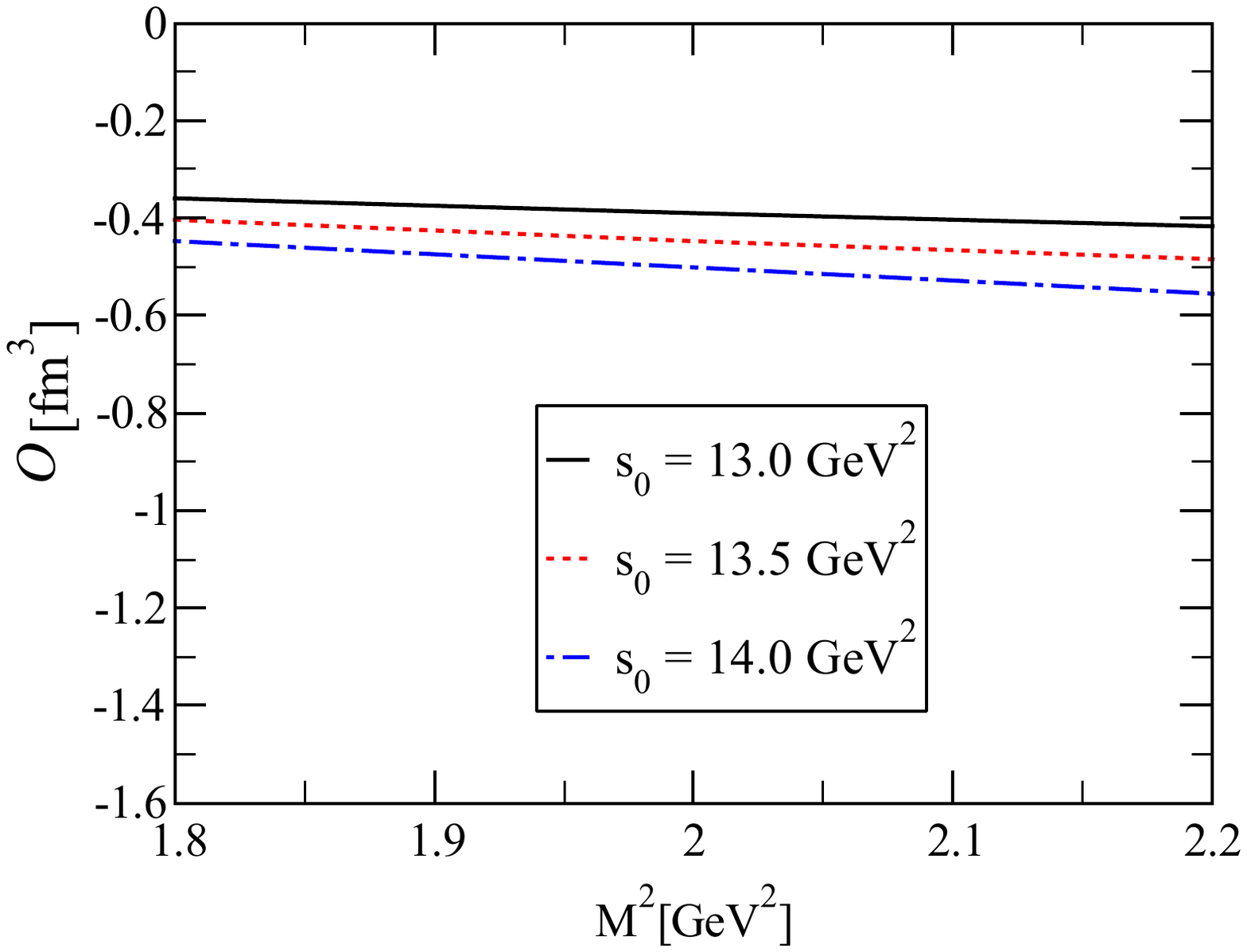}}~~~~~~~
 \subfloat[]{\includegraphics[width=0.37\textwidth]{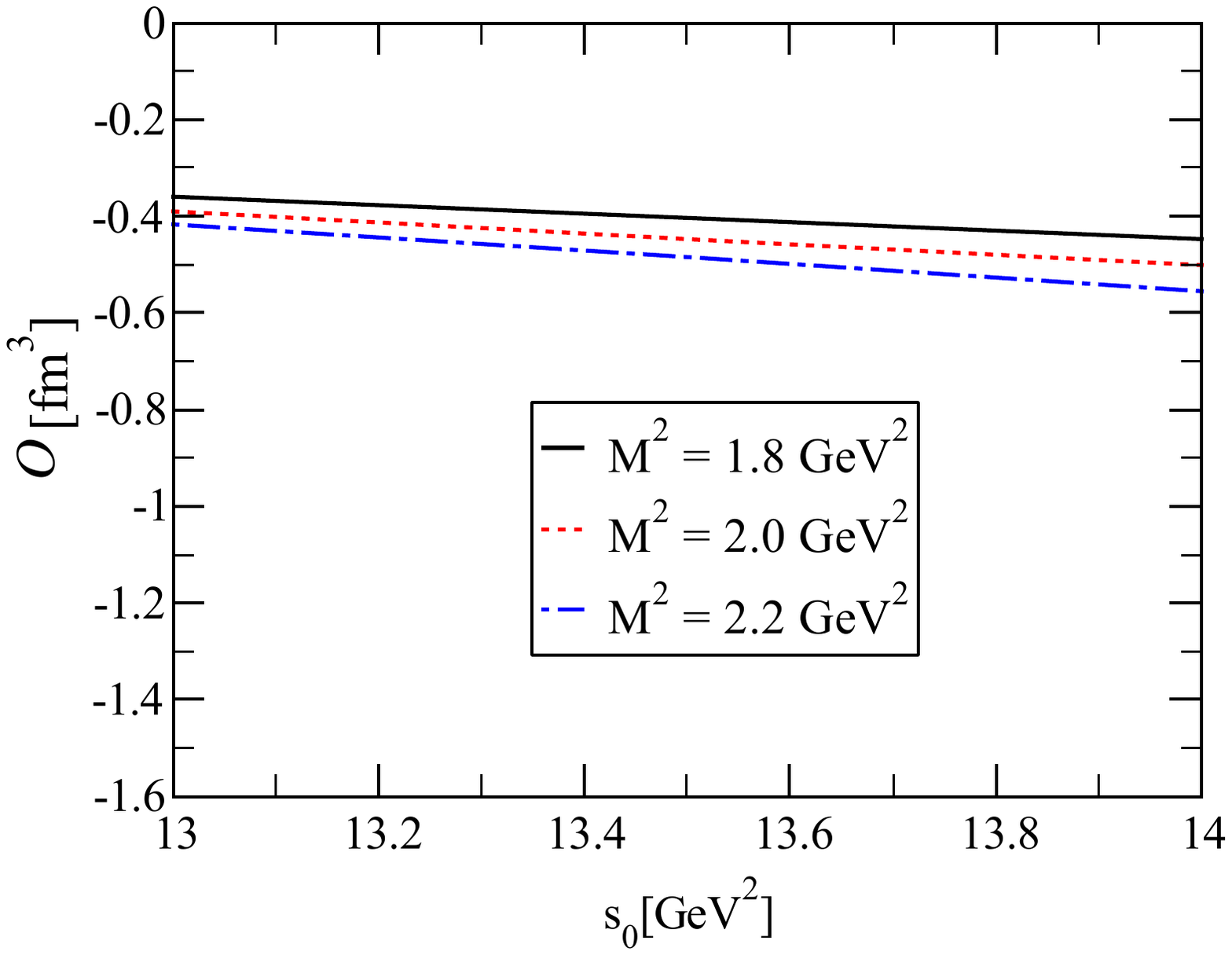}}
 \caption{ The the magnetic dipole, electric quadrupole ($\times 10^{-3}$), and magnetic octupole moments ($\times 10^{-4}$) of the $\Xi_c^*\bar K$ state as  a function of $\rm{M^2}$ (left-panel),  and $\rm{s_0}$ (right-panel).}
 \label{Msqfig}
  \end{figure}
   \end{widetext}

In addition to the parameters previously outlined, the methodology employed in this study incorporates two additional parameters: the Borel mass parameter, denoted by $\rm{M^2}$, and the continuum threshold, designated by $\rm{s_0}$. In an ideal scenario, these two parameters are expected to exhibit minimal bias on the outcomes. In order to achieve this, it is necessary to identify the region in which the variation of the results regarding the aforementioned parameters should be minimal. This region is designated as the working region.  The working regions are prescribed by the method employed and are determined by two constraints, namely pole contribution (PC) and OPE convergence (CVG). Under the established methodology, the CVG is required to exhibit sufficient smallness in order to ensure convergence of the OPE. In contrast, the PC is expected to be sufficiently large so that the efficiency of the single-pole scheme can be enhanced.  To characterize these constraints, it is useful to use the following expressions:
\begin{align}
 \mbox{PC} &=\frac{\rho_i (\rm{M^2},\rm{s_0})}{\rho_i (\rm{M^2},\infty)},\\
 \mbox{CVG} &=\frac{\rho_i^{\mbox{DimN}} (\rm{M^2},\rm{s_0})}{\rho_i (\rm{M^2},\rm{s_0})} ,
 \end{align}
 where $\rho_i^{\mbox{DimN}} (\rm{M^2},\rm{s_0})$ represent the highest dimensional terms in the  OPE of the $\rho_i (\rm{M^2},\rm{s_0})$. As can be seen in the equations given in the appendix, the highest dimensional terms are dimension 9  and 10. Consequently, the CVG analysis has been performed by considering the DimN expression in the form of $\rm{Dim(9+10)}$.  
 As a result of these analyses, the PC and CVG restrictions are met in the region 
 \begin{align*}
\rm{s_0}\in [13.0, 14.0]~\mbox{GeV}^2, ~~~~ 
\rm{M^2} \in [1.8, 2.2]~\mbox{GeV}^2,
\end{align*}
for the $\Xi_c^*\bar K$ state. Our computations indicate that considering these working intervals for the $\rm{M^2}$, the magnetic dipole and higher multipole moments of the $\Xi_c^*\bar K$ state PC vary within the interval $25.2\%\leq \rm{PC} \leq 46.4\%$ for the $\Xi_c^*\bar K$ state. 
Analyzing the CVG, one sees that the contribution of the higher dimensional terms in the OPE is less than $1.5 \%$ of the total and the series shows good convergence for the $\Xi_c^*\bar K$ state.  To illustrate, Fig. \ref{Msqfig} depicts the variation of the predicted magnetic dipole, electric quadrupole, and magnetic octupole moments of the $\Xi_c^*\bar K$ state with $\rm{M^2}$ and $\rm{s_0}$. As illustrated in the accompanying figure, the magnetic dipole, electric quadrupole, and magnetic octupole moments of this state exhibit a slight dependence on the parameters $\rm{M^2}$ and $\rm{s_0}$. 

Once all the pertinent parameters have been established, we are in a position to determine the numerical values associated with the magnetic dipole, electric quadrupole, and magnetic octupole moments of the $\Xi_c^{*} \bar K$ state. The extracted outcomes are given as  
  \begin{align}
   \mu_{\Xi_c^*\bar K} &= 0.15^{+0.04}_{-0.03}\,\mu_N,\\
   \mathcal{Q}_{\Xi_c^*\bar K} &= (-0.93^{+0.22}_{-0.17})\times 10^{-3}\,\rm{fm^2}, \\
   \mathcal{O}_{\Xi_c^*\bar K} &= (-0.45^{+0.10}_{-0.09})\times 10^{-4}\,\rm{fm^3}.
  \end{align}
The errors observed in the results can be attributed to uncertainties in the input parameters together with errors in the prediction of the working intervals of the auxiliary parameters.

We highlight our predictions as follows:

\begin{itemize}

 \item The resultant magnetic dipole, electric quadrupole, and magnetic octupole moments are governed by the short-distance interactions.  The short-distance contributions account for approximately 70$\%$ of the overall results, while the remaining part is comprised of long-distance contributions.
 
 \item We are able to check the individual quark contributions to the multipole moments with the help of the corresponding quark charges ($e_u$, $e_d$, $e_s$, and $e_c$). Our predictions indicate that the c-quark dominates over the light quarks and governs the multipole moments. Looking at the equations given in the appendix, it can be seen that the short-distance interactions are only proportional to $e_c$. This means that the contribution of light quarks in short-distance interactions cancel each other out and reduce their overall effects. 
 
 \item The order of the magnetic dipole moment may be employed to provide an indication of the experimental accessibility of the pertinent physical quantities. The order of magnitude of the results indicates that they may be measured in future experiments. 
 
 \item When the electric quadrupole and magnetic octupole moments results achieved are examined, the results are found to be small but non-zero, indicating a non-symmetrical charge distribution. 
 
 \item From the sign of the electric quadrupole moment result, it can be concluded that the geometric shape of the $\Xi_c^*\bar K$ state is oblate.
  
 \item In~\cite{Ozdem:2024pyb}, magnetic dipole moments were obtained for the $\Omega_c$ states by considering $[ss][dc][\bar d]$, $[sd][sc][\bar d]$ and $[ds][sc][\bar d]$ diquark-diquark-antiquark  configurations with quantum numbers $J^P = \frac{3}{2}^-$. The results obtained are as $-0.058^{+0.032}_{-0.031} \mu_N $, $-0.37^{+0.05}_{-0.07} \mu_N$, and $-0.37^{+0.05}_{-0.07}\mu_N $, respectively. The outcomes observed for the magnetic dipole moments obtained through the two configurations exhibited a notable discrepancy, not only in magnitude but also in sign. The measurement of magnetic dipole moments through experimental means may offer insight into the internal structure of these states.
\end{itemize}

\section{Summary}\label{secIV}

In conclusion, this paper examines the magnetic dipole, electric quadrupole, and magnetic octupole moments of the $\Xi_c^{*}\bar K$ state with spin-parity $J^P = \frac{3}{2}^-$ within the framework of the QCD light-cone sum rule, wherein this state is modeled as a molecular configuration. A quantitative analysis of the predicted results reveals that the magnetic dipole moment is sufficiently large to be experimentally measurable, while the electric quadrupole and magnetic octupole moments are achieved as a small but non-zero value corresponding to the oblate charge distribution. It is also recommended that our predictions be subjected to further scrutiny through the application of other phenomenological methodologies. The findings of this study on the magnetic dipole, electric quadrupole, and magnetic octupole moments of the $\Xi_c^{*} \bar K$ state may prove to be a valuable resource for future experimental investigations of exotic states. In addition to the electromagnetic properties, the decay channels, decay modes, and branching ratios play a crucial role in elucidating the inner structure of the particle in question. The nature of the $\Omega_c$ states remains uncertain and is the subject of ongoing debate. In order to gain further insight, it is necessary for experimental collaborations to undertake further efforts to explore these states, with a particular focus on determining their spin and parity, as well as their internal structure.

\begin{widetext}
\section*{Appendix: Explicit forms of the sum rules for $F_i(0)$}
 
This appendix presents the explicit expressions of the analytical expressions derived for the magnetic dipole, electric quadrupole, and magnetic octupole moments of the $\Xi_c^* \bar K$ state. The magnetic dipole, electric quadrupole, and magnetic octupole moments in terms of the $F_i(0)$ form factors are written as follows:
 \begin{eqnarray}
\mu_{{\Xi_c^* \bar K}}&=&\frac{e}{2m_{{\Xi_c^* \bar K}}}\rho_1(\rm{M^2},\rm{s_0}),\nonumber\\
Q_{{\Xi_c^* \bar K}}&=&\frac{e}{m_{{\Xi_c^* \bar K}}^2}\rho_2(\rm{M^2},\rm{s_0}),\nonumber\\
O_{{\Xi_c^* \bar K}}&=&\frac{e}{2m_{{\Xi_c^* \bar K}}^3}\rho_3(\rm{M^2},\rm{s_0}),
\end{eqnarray}
with 
\begin{align}
 \rho_1(\rm{M^2},\rm{s_0})&= F_{1}(0)+F_{2}(0),\nonumber\\
 \rho_2(\rm{M^2},\rm{s_0})&= F_{1}(0)-\frac{1}{2}F_{3}(0),\nonumber\\
 \rho_3(\rm{M^2},\rm{s_0})&= F_{1}(0)+F_{2}(0)-\frac{1}{2}[F_{3}(0)+F_{4}(0)],
\end{align}

where
\begin{align}
 F_1(0)&=  \frac{e^{\frac{m^2_{\Xi_c^*\bar K}}{\rm{M^2}}}}{\,\lambda^2_{\Xi_c^*\bar K}}\Bigg\{ -\frac {11 e_c } {2^{20} \times 3^2 \times 5^2 \times m_c^2 \times \pi^8}\Big[ 
   m_c^{16} I[-8, 5] - 15 m_c^{12} I[-6, 5] - 40 m_c^{10} I[-5, 5]- 
    45 m_c^8 I[-4, 5] \nonumber\\
   &  - 24 m_c^6 I[-3, 5]- 5 m_c^4 I[-2, 5] - 
    128 I[0, 5]\Big] \nonumber\\
    &
   +\frac {\langle g_s^2 G^2 \rangle \langle\bar qq \rangle \langle\bar ss \rangle} {2^{17} \times 3^4 \times m_c^2 \times \pi^4} \Big[ -16 (e_s - 
         e_u)  (m_c^4 I[-2, 0] - I[0, 0])\mathbb A[u_ 0] + 
      7 (19 e_s + 20 e_u)  (m_c^4 I[-2, 0] - 
         I[0, 0]) I_ 3[\mathcal S] \nonumber\\
    &+ 
      8 e_d  (m_c^4 I[-2, 0] - I[0, 0]) I_ 4[\mathcal S] + 
      16 \chi (e_s - e_u) \Big (4 m_c^6 I[-3, 1] + 
          m_c^4 (-m_ 0^2 I[-2, 0] + 4 I[-2, 1]) \nonumber\\
    & + 
          m_ 0^2 I[0, 0]\Big) \varphi_\gamma[u_ 0]\Big]\nonumber\\
             &  -\frac {\langle g_s^2 G^2 \rangle \langle\bar qq \rangle^2} {2^{18} \times 3^4 \times m_c^2 \times \pi^4} \Big[ \big (16 (e_d - e_u) \mathbb A[u_ 0] + 
        16 e_u I_ 3[\mathcal S] + 
        13 e_d I_ 4[\mathcal S]\big) \big (m_c^4 I[-2, 0] - 
       I[0, 0]\big) \nonumber\\
    & - 
    16 \chi (e_d - e_u) \Big (4 m_c^6 I[-3, 1] + 
        m_c^4 (-m_ 0^2 I[-2, 0] + 4 I[-2, 1]) + 
        m_ 0^2 I[0, 0]\Big) \varphi_\gamma[u_ 0]\Big]\nonumber\\
      &  -\frac {\langle g_s^2 G^2 \rangle \langle\bar ss \rangle^2} {2^{14} \times 3^4 \times m_c^2 \times \pi^4}  (m_c^4 I[-2, 0] - I[0, 0])I_3[\mathcal S]\nonumber\\
        &-\frac {m_s\langle\bar qq \rangle \langle\bar ss \rangle^2} {2^{10} \times 3^2 \times m_c^2 \times \pi^2} \Big[ 
   e_d I_ 4[\mathcal S] (-I[0, 0] + 
       m_c^6 I[3, 
         0]) + \Big (5 e_u \big (-m_c^4 I[-2, 0] + I[0, 0]\big) + 
       e_s \big (-5 m_c^4 I[-2, 0] \nonumber\\
       &+ 4 I[0, 0] + 
           m_c^6 I[3, 0]\big)\Big) I_ 3[\mathcal S]\Big]\nonumber\\
           &-\frac {m_s \langle\bar qq \rangle^2 \langle\bar ss \rangle} {2^{10} \times 3^2 \times m_c^2 \times \pi^2} \Big[(10 e_s + 23 e_u) I_3[\mathcal S] - e_d I_4[\mathcal S]) (m_c^4 I[-2, 0] - I[0, 0])\Big]\nonumber\\
                     & -\frac {m_s \langle g_s^2 G^2 \rangle \langle\bar qq \rangle} {2^{17} \times 3^4 \times m_c^2 \times \pi^6} \Big[ 
   3 m_c^4 \Big (-\big (4 e_u \mathbb A[u_ 0] + 
           35 e_u I_ 3[\mathcal S] + 
           2 e_d I_ 4[\mathcal S]\big) \big (m_c^2 I[-3, 1] + 
          I[-2, 1]\big) + 
       4 \chi e_u \big (m_c^4 I[-4, 2] \nonumber\\
       &- 2 m_c^2 I[-3, 2] + 
           I[-2, 2]\big) \varphi_\gamma[u_ 0]\Big) - 
    16 f_ {3\gamma} \pi^2 \Big (e_d m_c^4 I[-2, 0] + 
        e_s m_c^4 I[-2, 0] - e_d I[0, 0] - 6 e_s I[0, 0] - 
        5 e_u I[0, 0] \nonumber\\
       & + 5 (e_s + e_u) m_c^6 I[3, 0]\Big) \psi^
       a[u_ 0]\Big]\nonumber\\
       &-\frac {m_s \langle g_s^2 G^2 \rangle \langle\bar ss \rangle} {2^{17} \times 3^4 \times m_c^2 \times \pi^6} \Big[ 
   6 e_s m_c^4 I_ 3[\mathcal S] (m_c^2 I[-3, 1] + I[-2, 1]) + 
    f_ {3\gamma}\pi^2 \Big (\big ((2 e_s + 35 e_u) I_ 1[\mathcal V] - 
            2 e_d I_ 2[\mathcal V]\big) \big (m_c^4 I[-2, 0]\nonumber\\
       & - 
           I[0, 0]\big) + 
        16 (e_d + 3 e_s + 3 e_u)\big (I[0, 0] - 
            m_c^6 I[3, 0]\big) \psi^a[u_ 0]\Big)\Big]
            \nonumber
             \end{align}
       
           \begin{align}
       & -\frac {\langle\bar qq \rangle^2} {2^{14} \times 3^2 \times m_c^2 \times \pi^4} \Big[ 
   3 e_u m_c^4  \Big (m_c^4 I[-4, 2] - 
       m_c^2 (m_ 0^2 I[-3, 1] + 2 I[-3, 2]) - m_ 0^2 I[-2, 1] + 
       I[-2, 2]\Big) I_ 3[\mathcal S] \nonumber\\
            & + 
    3 e_d m_c^4  \Big (m_c^4 I[-4, 2] - 
       m_c^2 (m_ 0^2 I[-3, 1] + 2 I[-3, 2]) - m_ 0^2 I[-2, 1] + 
       I[-2, 2]\Big) I_ 4[\mathcal S] - 
    8 e_s f_ {3\gamma} \pi^2 \Big (2 m_c^6 I[-3, 1] \nonumber\\
           &+ 
       m_c^4 (-m_ 0^2 I[-2, 0] + 2 I[-2, 1]) + 
       m_ 0^2 I[0, 0]\Big) I_ 1[\mathcal V] \Big]\nonumber\\
            &-\frac {\langle\bar ss \rangle^2} {2^{14} \times 3^2 \times m_c^2 \times \pi^4} \Big[-4 f_ {3\gamma} \pi^2 \big (e_u I_ 1[\mathcal V] - e_d I_ 2[\mathcal V]\big) \Big (2 m_c^6 I[-3, 1] \nonumber\\
       &+ 
       m_c^4 (-m_ 0^2 I[-2, 0] + 2 I[-2, 1]) + m_ 0^2 I[0, 0]\Big) + 
    e_s  \Big (m_c^4 \big (2 m_c^6 I[-5, 2] + 
          3 m_ 0^2 m_c^4 I[-4, 1] - 6 m_c^2 I[-3, 2]\nonumber\\
       & - 
          3 m_ 0^2 I[-2, 1] + 4 I[-2, 2]\big) + 
       12 m_ 0^2 I[0, 1]\Big) I_ 3[\mathcal S] \Big]\nonumber\\
       &-\frac {\langle\bar qq \rangle \langle\bar ss \rangle} {2^{14} \times 3^2 \times m_c^2 \times \pi^4} \Bigg[ 8 f_ {3\gamma} \pi^2\big (4 e_s I_ 1[\mathcal V] + 
     5 e_u I_ 1[\mathcal V] + 
     e_d I_ 2[\mathcal V]\big) \Big (m_c^4 \big (2 m_c^2 I[-3, 1] - 
       m_ 0^2 I[-2, 0] + 2 I[-2, 1]\big) \nonumber\\
       &+ m_ 0^2 I[0, 0]\Big) + 
 e_d \Big (m_c^4 (2 m_c^6 I[-5, 2] + 3 m_ 0^2 m_c^4 I[-4, 1] - 
       6 m_c^2 I[-3, 2] - 3 m_ 0^2 I[-2, 1] + 4 I[-2, 2]) \nonumber\\
   %
     & + 
    12 m_ 0^2 I[0, 
       1]\Big) I_ 4[\mathcal S] + \Bigg (m_c^4 \Big (e_s \big (2 \
m_c^6 I[-5, 2] + 3 m_c^4 (m_ 0^2 I[-4, 1] - 10 I[-4, 2]) + 
          6 m_c^2 (5 m_ 0^2 I[-3, 1] + 9 I[-3, 2]) \nonumber\\
       & + 
          27 m_ 0^2 I[-2, 1] - 26 I[-2, 2]\big) + 
       30 e_u \big (-m_c^4 I[-4, 2] + 
           m_c^2 (m_ 0^2 I[-3, 1] + 2 I[-3, 2]) + m_ 0^2 I[-2, 1] - 
           I[-2, 2]\big)\Big) \nonumber\\
       &+ 
    12 e_s m_ 0^2 I[0, 1]\Bigg) I_ 3[\mathcal S] \Bigg]\nonumber\\
    &-\frac {m_c^2 \langle g_s^2 G^2 \rangle f_ {3\gamma}} {2^{23} \times 3^4  \times \pi^6}\Big[ 9 \big((94 e_s + 88 e_u) I_1[\mathcal V] - e_d I_2[\mathcal V]\big) (m_c^4 I[-4, 2] - 
    2 m_c^2 I[-3, 2] + I[-2, 2]) - 
 32 (e_d + 14 e_s + 9 e_u) \nonumber\\
 & \times \big(m_c^6 I[-5, 2] - 3 m_c^2 I[-3, 2] + 
    2 I[-2, 2]\big) \psi^a[u_0]\Big]\nonumber\\
    &- \frac{m_s \langle \bar qq \rangle}{2^{15} \times 3^2  \times m_c^2  \times \pi^6} \Big[-6 f_ {3\gamma} m_c^4 \pi^2 \big ((4 e_s + 5 e_u) I_ 1[\mathcal V] + 
     e_d I_ 2[\mathcal V]\big) \Big (m_c^4 I[-4, 2] - 
    m_c^2 (m_ 0^2 I[-3, 1] + 2 I[-3, 2])  \nonumber\\
 & - m_ 0^2 I[-2, 1] + 
    I[-2, 2]\Big) + 
 5 e_u  \Big (m_c^{10} I[-5, 3] + 3 m_c^8 I[-4, 3] + 
    3 m_c^6 I[-3, 3] + m_c^4 I[-2, 3] + 
    8 I[0, 3]\Big) I_ 3[\mathcal S]  \nonumber\\
 &- 
 e_d  \Big (m_c^{10} I[-5, 3]+ 3 m_c^8 I[-4, 3] + 3 m_c^6 I[-3, 3] + 
    m_c^4 I[-2, 3] + 8 I[0, 3]\Big) I_ 4[\mathcal S] \Big]
    \nonumber\\
    &
    + \frac{m_s \langle \bar ss \rangle}{2^{15} \times 3^2  \times m_c^2 \times  \pi^6} \Big[f_ {3\gamma} m_c^4 \pi^2\Bigg ( \Big (e_u (-39 m_c^4 I[-4, 2] + 
          m_c^2 \big (28 m_ 0^2 I[-3, 1] + 78 I[-3, 2]\big) + 
          28 m_ 0^2 I[-2, 1]  \nonumber\\
    &- 39 I[-2, 2]) + 
       5 e_s (-3 m_c^4 I[-4, 2] + 
           2 m_c^2 \big (m_ 0^2 I[-3, 1] + 3 I[-3, 2]\big) + 
           2 m_ 0^2 I[-2, 1] - 3 I[-2, 2])\Big) I_ 1[\mathcal V]  \nonumber\\
    &+ 
    e_d \Big (3 m_c^4 I[-4, 2]- 
       2 m_c^2 \big (2 m_ 0^2 I[-3, 1] + 3 I[-3, 2]\big) - 
       4 m_ 0^2 I[-2, 1] + 3 I[-2, 2]\Big) I_ 2[\mathcal V] \Bigg) + 
 e_s  \Big (m_c^{10} I[-5, 3]  \nonumber\\
    &+ 3 m_c^8 I[-4, 3] + 3 m_c^6 I[-3, 3]+ 
    m_c^4 I[-2, 3] + 8 I[0, 3]\Big) I_ 3[\mathcal S]   \Big]
           \Bigg\},
\end{align}

\begin{align}
 F_2(0)&= m_{\Xi_c^*\bar K}\frac{e^{\frac{m^2_{\Xi_c^*\bar K}}{\rm{M^2}}}}{\,\lambda^2_{\Xi_c^*\bar K}}\Bigg\{\frac {11 e_c } {2^{20} \times 3^2 \times 5^2 \times m_c^2 \times \pi^8}\Big[ 
   m_c^{14} I[-7, 5] - 15 m_c^{10} I[-6, 5] - 40 m_c^8 I[-4, 5] - 
    45 m_c^6 I[-3, 5]  \nonumber\\
   &- 24 m_c^4 I[-2, 5]- 5 m_c^2 I[-1, 5] - 
    128 I[0, 5]\Big]\nonumber\\
        &+\frac {\langle g_s^2 G^2 \rangle \langle\bar qq \rangle \langle\bar ss \rangle} {2^{18} \times 3^4 \times m_c \times \pi^4} \Big[16 \chi (e_s - e_u) \Big (2 m_c^6 I[-3, 1] + 4 m_c^4 I[-2, 1] + 
    8 I[0, 1] + 
    m_c^2 (2 I[-1, 1] - 
        m_ 0^2 I[1, 0])\Big) I_ 5[\varphi_\gamma] \nonumber\\
    &+ (9 e_s + 
    20 e_u) m_c^2 (I[-1, 0] - 
    m_c^2 I[2, 0]) I_ 1[\mathcal S] + -7 (19 e_s + 
    20 e_u)  (m_c^2 I[-1, 0] - I[0, 0]) I_ 3[\mathcal S] + 
 2 e_d m_c^2  (I[-1, 0] \nonumber
        \end{align}

\begin{align}
&- m_c^2 I[2, 0]) I_ 2[\mathcal S] + 
 8 \Big (e_d  (-m_c^2 I[-1, 0] + I[0, 0]) I_ 4[\mathcal S] - 
    2 (e_s - e_u)  (I[0, 0] - m_c^4 I[2, 0])\mathbb A[u_ 0] \nonumber\\
    &+ 
    m_c^2 \big ((-e_s + e_u) I_ 5[\mathcal A] I[1, 
         0] - (e_d + e_s + 2 e_u) I_ 6[
         h_\gamma] (I[1, 0] - 2 m_c^2 I[2, 0] + m_c^4 I[3, 0])\big) \nonumber\\
    &+ 
    2 \chi (e_s - 
       e_u) (2 m_c^6 I[-3, 1] + 2 m_c^2 I[-1, 1] - m_ 0^2 I[0, 0] + 
        8 I[0, 1] + m_ 0^2 m_c^4 I[2, 0]) \varphi_\gamma[u_ 0]\Big)
    \Big]\nonumber\\
     & +\frac {\langle g_s^2 G^2 \rangle \langle\bar qq \rangle^2} {2^{18} \times 3^4 \times m_c \times \pi^4} \Big[ -8 e_d m_c^2 I_ 2[\mathcal S] I[-1, 0] + 
 16 e_u m_c^2 I_ 3[\mathcal S] I[-1, 0] + 
 13 e_d m_c^2 I_ 4[\mathcal S] I[-1, 0] \nonumber\\
    &+16 (e_u-e_d) \mathbb A[u_ 0] I[0, 0] - 
 16 e_u I_ 3[\mathcal S] I[0, 0] - 13 e_d I_ 4[\mathcal S] I[0, 0] - 
 8 e_d m_c^2 I_ 5[\mathbb A] I[1, 0] + 
 8 e_u m_c^2 I_ 5[\mathbb A] I[1, 0] \nonumber\\
    & - 
 16 e_u m_c^2 I_ 6[h_\gamma] I[1, 0] + 
 8 \chi (e_d - e_u) I_ 5[\varphi_\gamma] \Big (2 m_c^6 I[-3, 1] + 
    4 m_c^4 I[-2, 1] + 8 I[0, 1] + 
    m_c^2 (2 I[-1, 1] - m_ 0^2 I[1, 0])\Big)\nonumber\\
    & + 
 16 e_d m_c^4 \mathbb A[u_ 0] I[2, 0] - 
 16 e_u m_c^4 \mathbb A[u_ 0] I[2, 0] + 
 8 e_d m_c^4 I_ 2[\mathcal S] I[2, 0] + 
 32 e_u m_c^4 I_ 6[h_\gamma] I[2, 0] + 
 4 e_u m_c^2 I_ 1[\mathcal S] (I[-1, 0] \nonumber\\
    &- m_c^2 I[2, 0]) - 
 16 e_u m_c^6 I_ 6[h_\gamma] I[3, 0] + 
 16 \chi (e_d - e_u) \Big (2 m_c^6 I[-3, 1] + 2 m_c^2 I[-1, 1] - 
     m_ 0^2 I[0, 0] + 8 I[0, 1] \nonumber\\
    &+ 
     m_ 0^2 m_c^4 I[2, 0]\Big) \varphi_\gamma[u_ 0]\Big]\nonumber\\
     &-\frac {e_s\,\langle g_s^2 G^2 \rangle \langle\bar ss \rangle^2} {2^{16} \times 3^4 \times m_c \times \pi^4} \Big[ 4 I_3[\mathcal S] (m_c^2 I[-1, 0] - I[0, 0]) + 
 I_1[\mathcal S] (-m_c^2 I[-1, 0] + m_c^4 I[2, 0]) + 
 4 m_c^2 I_6[h_\gamma] \big(I[1, 0]\nonumber\\
    & - 2 m_c^2 I[2, 0] + m_c^4 I[3, 0]\big)
     \Big]\nonumber\\
     &+\frac {m_s \langle \bar qq \rangle \langle\bar ss \rangle^2} {2^{10} \times 3^2 \times m_c \times \pi^2} \Big[e_d I_4[\mathcal S] (-I[0, 0] + m_c^4 I[2, 0]) + 
 I_3[\mathcal S] \Big(5 e_u (-m_c^2 I[-1, 0] + I[0, 0]) + 
    e_s (-5 m_c^2 I[-1, 0] \nonumber\\
    &+ 4 I[0, 0] + m_c^4 I[2, 0])\Big)\Big]\nonumber\\
    &+\frac {m_s \langle \bar ss \rangle \langle\bar qq \rangle^2} {2^{10} \times 3^2 \times m_c \times \pi^2} \Big[((10 e_s + 23 e_u) I_3[\mathcal S] - e_d I_4[\mathcal S]) (m_c^2 I[-1, 0] - I[0, 0])\Big]\nonumber\\
    & -\frac {m_s \langle g_s^2 G^2 \rangle \langle\bar qq \rangle} {2^{19} \times 3^4 \times m_c \times \pi^6} \Big[ 24 e_u m_c^6 \mathbb A[u_ 0] I[-3, 1] + 
 15 e_u m_c^6 I_1[\mathcal S] I[-3, 1] + 
 3 e_d m_c^6 I_ 2[\mathcal S] I[-3, 1] + 
 12 e_u m_c^6 I_ 5[\mathbb A] I[-3, 1] \nonumber\\
 &+ 
 30 e_u m_c^4 I_ 1[\mathcal S] I[-2, 1] + 
 6 e_d m_c^4 I_ 2[\mathcal S] I[-2, 1] - 
 420 e_u m_c^4 I_ 3[\mathcal S] I[-2, 1] - 
 24 e_d m_c^4 I_ 4[\mathcal S] I[-2, 1] + 
 24 e_u m_c^4 I_ 5[\mathbb A] I[-2, 1] \nonumber\\
 &+ 
 24 e_u m_c^2 \mathbb A[u_ 0] I[-1, 1] + 
 15 e_u m_c^2 I_ 1[\mathcal S] I[-1, 1] + 
 3 e_d m_c^2 I_ 2[\mathcal S] I[-1, 1] + 
 12 e_u m_c^2 I_5[\mathbb A] I[-1, 1] + 
 8 \chi e_u m_c^2 I_ 5[
   \varphi_\gamma]\nonumber\\
 & \times  (m_c^6 I[-4, 2] - 3 m_c^4 I[-3, 2] + 
    3 m_c^2 I[-2, 2] - I[-1, 2]) + 96 e_u \mathbb A[u_ 0] I[0, 1] + 
 60 e_u I_ 1[\mathcal S] I[0, 1] + 12 e_d I_ 2[\mathcal S] I[0, 1]\nonumber\\
 & + 
 48 e_u I_ 5[\mathbb A] I[0, 1] - 
 8 (e_d + 2 e_u) I_ 6[h_\gamma] \Big (m_c^8 I[-4, 1] + 
    3 m_c^6 I[-3, 1] + 3 m_c^4 I[-2, 1] + m_c^2 I[-1, 1] + 
    8 I[0, 1]\Big) \nonumber\\
 &+ 
 32 (e_d + 6 e_s + 5 e_u) f_ {3\gamma} m_c^2 \pi^2 I_ 5[\psi^a] I[1, 
   0] + 16 \chi e_u m_c^2 (m_c^6 I[-4, 2] - 3 m_c^2 I[-2, 2] + 
     2 I[-1, 2]) \varphi_\gamma[u_ 0] \nonumber\\
 &- 
 64 f_ {3\gamma} \pi^2 \Big (-5 e_s m_c^2 I[-1, 0] - 
     5 e_u m_c^2 I[-1, 0] - e_d I[0, 0] - 6 e_s I[0, 0] - 
     5 e_u I[0, 0] + (e_d + 11 e_s + 10 e_u)\nonumber\\
 & \times  m_c^4 I[2, 0]\big) \psi^a[u_ 0]\Big]\nonumber\\
    &
    +\frac {m_s \langle g_s^2 G^2 \rangle \langle\bar ss \rangle} {2^{19} \times 3^4 \times m_c \times \pi^6} \Bigg[24 e_s m_c^4 I_3[\mathcal S] I[-2, 1] - 
 3 e_s I_ 1[\mathcal S] \Big (m_c^6 I[-3, 1] + 2 m_c^4 I[-2, 1] + 
    m_c^2 I[-1, 1] + 4 I[0, 1]\Big) \nonumber\\
    &+ 
 8 e_s I_ 6[h_\gamma] \Big (m_c^8 I[-4, 1] + 3 m_c^6 I[-3, 1] + 
    3 m_c^4 I[-2, 1] + m_c^2 I[-1, 1] + 8 I[0, 1]\Big) + 
 2 f_ {3\gamma} \pi^2  \Bigg (e_d I_ 2[\mathcal V] (-7 m_c^2 I[-1, 
         0] \nonumber\\
    &+ 4 I[0, 0] + 3 m_c^4 I[2, 0]) + 
    I_ 1[\mathcal V] \Big (-e_s m_c^2 I[-1, 0] + 
       65 e_u m_c^2 I[-1, 0] - 4 e_s I[0, 0] - 70 e_u I[0, 0] + 
       5 (e_s + e_u) m_c^4 I[2, 0]\Big)\nonumber\\
    & + 
    16 (e_d + 3 e_s + 3 e_u) \Big (m_c^2 I_ 5[\psi^a] I[1, 0] + 
        2 (m_c^2 I[-1, 0] + I[0, 0] - 2 m_c^4 I[2, 0]) \psi^
           a[u_ 0]\Big)\Bigg)
    \Bigg]\nonumber\\
    & + \frac { \langle\bar qq \rangle^2} {2^{14} \times 3^2 \times m_c \times \pi^4} \Big[3 e_u m_c^2  \Big (m_c^4 I[-3, 2] - 
    m_c^2 (m_ 0^2 I[-2, 1] + 2 I[-2, 2]) + 
    I[-1, 2]\Big) I_ 3[\mathcal S] + 
 3 e_d m_c^2 \Big (m_c^4 I[-3, 2] \nonumber\\
    &- 
    m_c^2 (m_ 0^2 I[-2, 1] + 2 I[-2, 2]) + 
    I[-1, 2]\Big) I_ 4[\mathcal S] - 
 8 e_s f_ {3\gamma}\pi^2  \Big (2 m_c^4 I[-2, 1] - 
    m_ 0^2 m_c^2 I[-1, 0] + m_ 0^2 I[0, 0]\Big) I_ 1[\mathcal V] \Big]
    \nonumber\\
    & + \frac { \langle\bar ss \rangle^2} {2^{14} \times 3^2 \times m_c \times \pi^4} \Big[-4 f_ {3\gamma}\pi^2  (e_u I_ 1[\mathcal V] - 
     e_d I_ 2[\mathcal V]) \Big (2 m_c^4 I[-2, 
      1] - m_ 0^2 m_c^2 I[-1, 0] + m_ 0^2 I[0, 0]\Big)\nonumber
      \end{align}

\begin{align}
    & + 
 e_s\Big (2 m_c^8 I[-4, 2] + 3 m_ 0^2 m_c^6 I[-3, 1] - 
    6 m_c^4 I[-2, 2] + m_c^2 (3 m_ 0^2 I[-1, 1] + 4 I[-1, 2]) + 
    12 m_ 0^2 I[0, 1]\Big) I_ 3[\mathcal S]  \Big]
    \nonumber\\
         & + \frac { \langle\bar qq \rangle \langle\bar ss \rangle} {2^{14} \times 3^2 \times m_c \times \pi^4} \Bigg[  8 f_ {3\gamma}\pi^2 \Big (4 e_s I_ 1[\mathcal V] + 
    5 e_u I_ 1[\mathcal V] + 
    e_d I_ 2[\mathcal V]\Big) (2 m_c^4 I[-2, 1] - 
    m_ 0^2 m_c^2 I[-1, 0] + m_ 0^2 I[0, 0]) \nonumber\\
          &+ 
 e_d \Big (m_c^2 (2 m_c^6 I[-4, 2] + 3 m_ 0^2 m_c^4 I[-3, 1] - 
       6 m_c^2 I[-2, 2] + 3 m_ 0^2 I[-1, 1] + 4 I[-1, 2]) + 
    12 m_ 0^2 I[0, 
       1]\Big) I_ 4[\mathcal S] \nonumber\\
    &+ \Bigg (m_c^2 \Big (e_s (2 m_c^6 \
I[-4, 2] + 3 m_c^4 (m_ 0^2 I[-3, 1] - 10 I[-3, 2]) + 
          6 m_c^2 (5 m_ 0^2 I[-2, 1] + 9 I[-2, 2]) + 
          3 m_ 0^2 I[-1, 1] \nonumber\\
    &- 26 I[-1, 2]) - 
       30 e_u (m_c^4 I[-3, 2] - m_c^2 (m_ 0^2 I[-2, 1] + 2 I[-2, 2]) +
            I[-1, 2])\Big) + 
    12 e_s m_ 0^2 I[0, 1]\Bigg) I_ 3[\mathcal S] \Bigg]
    \nonumber\\
        &-\frac {m_c \langle g_s^2 G^2 \rangle f_ {3\gamma}} {2^{23} \times 3^4  \times \pi^6}\Big[e_d \Big (34 m_c^6 I[-4, 2] - 111 m_c^4 I[-3, 2] + 
    120 m_c^2 I[-2, 2] - 43 I[-1, 2]\Big) I_ 2[\mathcal V] + 
 3 \Big ((9 e_s + 8 e_u) m_c^6 I[-4, 2] \nonumber\\
    &+ 
    15 (17 e_s + 16 e_u) m_c^4 I[-3, 2] - 
    3 (179 e_s + 168 e_u) m_c^2 I[-2, 
      2] + (273 e_s + 256 e_u) I[-1, 2]\Big) I_ 1[\mathcal V] + 
 16 (e_d + 14 e_s + 
     9 e_u) \nonumber\\
    & \times \Big ( (-m_c^6 I[-4, 2] + 3 m_c^4 I[-3, 2] - 
       3 m_c^2 I[-2, 2] + I[-1, 2]) I_ 5[\psi^a] - 
    2 (2 m_c^6 I[-4, 2] - 3 m_c^4 I[-3, 2] + I[-1, 2]) \psi^
       a[u_ 0]\Big)\Big]\nonumber\\
    &+ \frac{m_s \langle \bar qq \rangle}{2^{15} \times 3^2  \times m_c  \times \pi^6} \Big[ -f_ {3\gamma}\pi^2 m_c^2  \big ((4 e_s + 5 e_u) I_ 1[\mathcal V] + 
     e_d I_ 2[\mathcal V]\big) \Big (m_c^4 I[-3, 2] - 
    m_c^2 (m_ 0^2 I[-2, 1] + 2 I[-2, 2]) + I[-1, 2]\Big) \nonumber\\
    &+ 
 5 e_u \Big (m_c^8 I[-4, 3] + 3 m_c^6 I[-3, 3] + 3 m_c^4 I[-2, 3] + 
    m_c^2 I[-1, 3] + 8 I[0, 3]\Big) I_ 3[\mathcal S] - 
 e_d \Big (m_c^8 I[-4, 3] + 3 m_c^6 I[-3, 3] \nonumber\\
    &+ 3 m_c^4 I[-2, 3] + 
    m_c^2 I[-1, 3] + 8 I[0, 3]\Big) I_ 4[\mathcal S] \Big]
    \nonumber\\
    &
    + \frac{m_s \langle \bar ss \rangle}{2^{15} \times 3^2  \times m_c \times  \pi^6} \Big[ f_ {3\gamma}\pi^2 m_c^2   \Big (m_c^2 (15 e_s m_c^2 I[-3, 2] + 
       39 e_u m_c^2 I[-3, 2] - 10 e_s m_ 0^2 I[-2, 1] - 
       28 e_u m_ 0^2 I[-2, 1] \nonumber\\
    &- 6 (5 e_s + 13 e_u) I[-2, 2]) + 
    3 (5 e_s + 13 e_u) I[-1, 2]\Big) I_ 1[\mathcal V] + 
 e_d f_ {3\gamma}\pi^2 \Big (2 m_c^8 I[-4, 2] + 
    m_c^6 (2 m_ 0^2 I[-3, 1] - 9 I[-3, 2]) \nonumber\\
    &+ 
    4 m_c^4 (2 m_ 0^2 I[-2, 1] + 3 I[-2, 2]) + 
    m_c^2 (2 m_ 0^2 I[-1, 1] - 5 I[-1, 2]) + 
    8 m_ 0^2 I[0, 1]\Big) I_ 2[\mathcal S] - 
 e_s  \Big (m_c^8 I[-4, 3] + 3 m_c^6 I[-3, 3] \nonumber\\
    &+ 3 m_c^4 I[-2, 3] + 
    m_c^2 I[-1, 3] + 8 I[0, 3]\Big) I_ 3[\mathcal S] \Big]
    \Bigg\},
\end{align}

\begin{align}
 F_3(0)&= 4m_{\Xi_c^*\bar K}\frac{e^{\frac{m^2_{\Xi_c^*\bar K}}{\rm{M^2}}}}{\,\lambda^2_{\Xi_c^*\bar K}}\Bigg\{\frac{11 e_c m_c}{2^{119} \times 3^2 \times 5 \times \pi^8}\Big[  m_c^{12} I[-7, 4] - 6 m_c^{10} I[-6, 4] + 15 m_c^8 I[-5, 4] - 
 20 m_c^6 I[-4, 4] + 15 m_c^4 \nonumber\\
 & \times I[-3, 4] - 6 m_c^2 I[-2, 4] + I[-1, 4]\Big]
 \nonumber\\
 &+\frac { m_c \chi  \langle g_s^2 G^2 \rangle \langle\bar qq \rangle \langle\bar ss \rangle} {2^{14} \times 3^4 \times \pi^4}(e_d + e_s + 2 e_u)  I[-1, 0] \varphi_\gamma[u_0]\nonumber\\
  &+\frac { m_s \langle g_s^2 G^2 \rangle \langle\bar qq \rangle } {2^{19} \times 3^4 \times \pi^6} \Big[ 3 m_c^2 \Big (2 (e_d + 2 e_u) \mathbb A[u_ 0] I[-1, 0] + 
    e_d \big (I[1, 0] - 2 m_c^2 I[2, 0] + 
       m_c^4 I[3, 0]\big) I_ 4[\mathcal S] \Big) \nonumber\\
 &- 
 8 \chi (e_d + 2 e_u) \Big (m_c^8 I[-4, 1] + 3 m_c^6 I[-3, 1] + 
     3 m_c^4 I[-2, 1] + m_c^2 I[-1, 1] + 
     8 I[0, 1]\Big) \varphi_\gamma[u_ 0]\Big]\nonumber\\
   &+\frac {e_s  m_s \langle g_s^2 G^2 \rangle \langle\bar ss \rangle } {2^{19} \times 3^4 \times \pi^6} \Big[ 3 m_c^2 \Big (2 \mathbb A[u_ 0] I[-1, 
      0] + \big (I[1, 0] - 2 m_c^2 I[2, 0] + 
       m_c^4 I[3, 0]\big) I_ 3[\mathcal S] \Big) - 
 8 \chi  \varphi_\gamma[u_ 0] \Big (m_c^8 I[-4, 1] \nonumber\\
 & + 3 m_c^6 I[-3, 1] + 3 m_c^4 I[-2, 1] + 
    m_c^2 I[-1, 1] + 8 I[0, 1]\Big)\Big]\nonumber\\
    &-\frac {  \langle\bar qq \rangle \langle\bar ss \rangle} {2^{15} \times m_c \times 3^2 \times \pi^4} \Big[ 320 (e_s + e_u) f_ {3\gamma}\pi^2 m_c^2 (I[-1, 0] - 
    m_c^2 I[2, 0]) I_ 3[\mathcal V] + 
 e_s  I_ 3[\mathcal S] \Big (4 m_c^8 I[-4, 1] + 32 I[0, 1]\nonumber\\
  & + 
    m_c^2 (4 I[-1, 1] - 3 m_ 0^2 I[1, 0]) + 
    6 m_c^4 (2 I[-2, 1] + m_ 0^2 I[2, 0]) + 
    3 m_c^6 (4 I[-3, 1] - m_ 0^2 I[3, 0])\Big)  \nonumber
 \end{align}

\begin{align}
 &+ 
 e_d  \Big (4 m_c^8 I[-4, 1] + 32 I[0, 1] + 
    m_c^2 (4 I[-1, 1] - 3 m_ 0^2 I[1, 0]) + 
    6 m_c^4 (2 I[-2, 1] + m_ 0^2 I[2, 0]) + 
    3 m_c^6 (4 I[-3, 1] \nonumber\\
 &- m_ 0^2 I[3, 0])\Big) I_ 4[\mathcal S] \Big]\nonumber\\
     &-\frac {\langle\bar ss \rangle^2} {2^{15} \times m_c \times 3^2 \times \pi^4} \Big[ 32 f_ {3\gamma}\pi^2 m_c^2 (e_u I_ 3[\mathcal V] - 
    e_d I_ 4[\mathcal V]) (-I[-1, 0] + m_c^2 I[2, 0]) + 
 e_s \Big (4 m_c^8 I[-4, 1] + 32 I[0, 1] \nonumber\\
 &+ 
    m_c^2 (4 I[-1, 1] - 3 m_ 0^2 I[1, 0]) + 
    6 m_c^4 (2 I[-2, 1] + m_ 0^2 I[2, 0]) + 
    3 m_c^6 (4 I[-3, 1] - m_ 0^2 I[3, 0])\Big) I_ 3[\mathcal S] \Big]\nonumber\\
      &+\frac { \langle g_s^2 G^2 \rangle f_ {3\gamma}} {2^{20} \times 3^4  \times m_c \times \pi^6}\Big[(7 e_s + 24 e_u) \Big (m_c^8 I[-4, 1] + 3 m_c^6 I[-3, 1] + 
    3 m_c^4 I[-2, 1] + m_c^2 I[-1, 1] + 
    8 I[0, 1]\Big) I_ 1[\mathcal V] \nonumber\\
     &- 
 23 e_d  \Big (m_c^8 I[-4, 1] + 3 m_c^6 I[-3, 1] + 3 m_c^4 I[-2, 1] + 
    m_c^2 I[-1, 1] + 8 I[0, 1]\Big) I_ 2[\mathcal V] + 
 2 e_d \Big (-4 m_c^8 I[-4, 1] \nonumber\\
 &+ 39 m_c^6 I[-3, 1] + 
    90 m_c^4 I[-2, 1] + 47 m_c^2 I[-1, 1] + 
    172 I[0, 1]\Big) I_ 4[\mathcal V] + 
 12  \Big ((e_s + 4 e_u) m_c^8 I[-4, 1] - 
    5 (23 e_s \nonumber\\
 %
  &+ 24 e_u) m_c^6 I[-3, 
      1] - (233 e_s + 252 e_u) m_c^4 I[-2, 
      1] - (117 e_s + 128 e_u) m_c^2 I[-1, 1] - 
    16 (29 e_s + 31 e_u) I[0, 1]\Big) I_ 3[\mathcal V] \nonumber\\
 &+ 
 32 (e_d + 2 e_s) \Big (m_c^8 I[-4, 1] + 3 m_c^6 I[-3, 1] + 
     3 m_c^4 I[-2, 1] + m_c^2 I[-1, 1] + 8 I[0, 1]) \psi^a[u_ 0]\Big]\nonumber\\
       &+\frac {5 m_s f_{3 \gamma}\langle \bar qq \rangle} {2^{15}  \times 3^2 \times m_c\times \pi^4} (e_s+e_u) \Big[\Big( 2 m_c^6 I[-3, 1] + m_c^2 (-m_0^2 I[-1, 0] + 2 I[-1, 1]) + 
 8 I[0, 1] + m_c^4 (4 I[-2, 1]  \nonumber\\
 &+ m_0^2 I[2, 0]) \Big)I_ 3[\mathcal V] \Big]\nonumber\\
        &-\frac { m_s f_{3 \gamma}\langle \bar ss \rangle} {2^{15}  \times 3^2 \times m_c\times \pi^4}  \Big[
        -4 e_d  \Big (9 m_c^6 I[-3, 1] + 
    m_c^2 (-4 m_ 0^2 I[-1, 0] + 9 I[-1, 1]) + 36 I[0, 1] + 
    2 m_c^4 (9 I[-2, 1] \nonumber\\
 &+ 2 m_ 0^2 I[2, 0])\Big) I_ 4[\mathcal V] + 
 4 \Big (60 e_s I[0, 1] + 156 e_u I[0, 1] + 
    5 e_s m_c^2 \big (3 m_c^4 I[-3, 1] - m_ 0^2 I[-1, 0] + 
       3 I[-1, 1] + m_c^2 (6 I[-2, 1] 
       \nonumber\\
 &+ m_ 0^2 I[2, 0])\big) + 
    e_u m_c^2 \big (39 m_c^4 I[-3, 1] - 14 m_ 0^2 I[-1, 0] + 
        39 I[-1, 1] + 
        2 m_c^2 (39 I[-2, 1] + 
            7 m_ 0^2 I[2, 0])\big)\Big) I_ 3[\mathcal V] 
            \nonumber\\
 &+ 
 e_d \big (2 m_c^8 I[-4, 1] + 16 I[0, 1] + 
    m_c^2 (2 I[-1, 1] - m_ 0^2 I[1, 0]) + 
    2 m_c^4 (3 I[-2, 1] + m_ 0^2 I[2, 0]) + 
    m_c^6 (6 I[-3, 1] \nonumber\\
 &- m_ 0^2 I[3, 0])\big) I_ 2[\mathcal V]
        \Big] 
    \Bigg\},
\end{align}
and, 
\begin{align}
 F_4(0)&= 4m^3_{\Xi_c^*\bar K}\frac{e^{\frac{m^2_{\Xi_c^*\bar K}}{\rm{M^2}}}}{\,\lambda^2_{\Xi_c^*\bar K}}
 \Bigg\{\frac{11 e_c }{2^{18} \times 3^3 \times 5 \times \pi^8}m_c^9 \big(m_c^2 I[-6, 3] + 5 I[-5, 3]\big)\nonumber\\
 &-\frac {m_c \langle g_s^2 G^2 \rangle f_ {3\gamma}} {2^{19} \times 3^4   \times \pi^6}\Big[
 3 \big(e_s + 4 e_u) I_3[\mathcal V] - 2 e_d I_4[\mathcal V] \big) \big(m_c^6 I[-4, 0] - 
   3 m_c^4 I[-3, 0] + 3 m_c^2 I[-2, 0] - I[-1, 0]\big)\Big]\nonumber\\
   &+\frac {f_{3 \gamma}\langle \bar qq \rangle} {2^{18}  \times 3^2 \times \pi^4} (e_s+e_u) \Big[\Big( 2 m_c^6 I[-4, 1] + m_c^2 (-m_0^2 I[-2, 0] + 2 I[-2, 1])  + m_c^4 (4 I[-3, 1]   \Big)I_ 3[\mathcal V] \Big]
    \Bigg\},
\end{align}
where $m_s$ and $m_c$ are the mass of the s-and  c-quark, respectively; $e_u$, $e_d$, $e_s$  and $e_c$ is the charge of the u-, d-, s- and c-quark, respectively; $\chi$ is the magnetic susceptibility of the quark condensate;  $m_0^2 =\langle \bar q\, g_s\, \sigma_{\alpha\beta}\, G^{\alpha\beta}\, q \rangle /\langle \bar qq \rangle $;   
$\langle \bar qq \rangle$, $\langle \bar ss \rangle$ and $\langle g_s^2 G^2\rangle$ are u/d-quark, s-quark and gluon condensates, respectively. 
The $I[n,m]$ and $I_i[\mathcal{F}]$ integrals are defined in the following manner:
\begin{align}
I[n,m]&= \int_{\mathcal{M}^2}^{\mathrm{s_0}} ds \int_{\mathcal{M}^2}^s dw \,w^n~(s-w)^m\, e^{-s/\mathrm{M^2}},\nonumber\\
I_1[\mathcal{F}]&=\int D_{\alpha_i} \int_0^1 dv~ \delta'(\alpha_ q +\bar v \alpha_g-u_0) \mathcal{F}(\alpha_{\bar q},\alpha_q,\alpha_g),\nonumber\\
I_2[\mathcal{F}]&=\int D_{\alpha_i} \int_0^1 dv~\delta'(\alpha_{\bar q}+ v \alpha_g-u_0) \mathcal{F}(\alpha_{\bar q},\alpha_q,\alpha_g),\nonumber
  \end{align}
 \begin{align}
I_3[\mathcal{F}]&=\int D_{\alpha_i} \int_0^1 dv~\delta(\alpha_ q +\bar v \alpha_g-u_0) \mathcal{F}(\alpha_{\bar q},\alpha_q,\alpha_g),\nonumber\\
I_4[\mathcal{F}]&=\int D_{\alpha_i} \int_0^1 dv~\delta(\alpha_{\bar q}+ v \alpha_g-u_0) \mathcal{F}(\alpha_{\bar q},\alpha_q,\alpha_g),\nonumber\\
   I_5[\mathcal{F}]&=\int_0^1 du~\delta'(u-u_0) \mathcal{F}(u),\nonumber\\
 I_6[\mathcal{F}]&=\int_0^1 du~ \mathcal{F}(u),
 \end{align}
 where $\mathcal{M} = m_c+2m_s $.
\end{widetext}
\bibliographystyle{elsarticle-num}
\bibliography{Xic0Kbarv1.bib}

\end{document}